\documentclass{elsart}
\usepackage{epsfig}
\usepackage{graphicx}
\usepackage{hyperref}
\usepackage{bbold}

\journal{Nuclear Instruments and Methods A}

\begin{document}

\vspace*{1cm}

\begin{frontmatter}
\title {Modelling and Measurement of Charge Transfer in 
        Multiple GEM Structures}

\author{M.~Killenberg},
\author{S.~Lotze},
\author{J.~Mnich}, 
\author{S.~Roth\thanksref{cor}},
\author{R.~Schulte}, 
\author{B.~Sobloher},
\author{W.~Struczinski\thanksref{dec}}, 
\author{M.~Tonutti}

\address{III.\ Physikalisches Institut, RWTH Aachen,\\
              D-52056 Aachen, Germany} 

\thanks[cor]{Corresponding author.
              {\it Phone:} +49--241--80--27298; 
              {\it Fax:} +49--241--80--22244;
              {\it Email:} {\tt roth@physik.rwth-aachen.de}}
\thanks[dec]{Deceased.}

\label{abstract}
\begin{abstract}
Measurements and numerical simulations on the charge transfer in Gas Electron Multiplier (GEM) foils are presented and their implications for the usage of GEM foils in Time Projection Chambers are discussed.  A small test chamber has been constructed and operated with up to three GEM foils.  The charge transfer parameters derived from the electrical currents monitored during the irradiation with an ${}^{55}$Fe source are compared with numerical simulations.
The performance in magnetic fields up to 2 T is also investigated.
\end{abstract}


\end{frontmatter}

\section{Introduction}
\label{introduction}
A Time Projection Chamber (TPC) is foreseen as the main tracker
of the detector for the TESLA linear collider~\cite{tdr}.
It should provide a momentum resolution of
$\delta (1/p_t) < 2 \cdot 10^{-4} (\mathrm{GeV/c})^{-1}$
to exploit especially the physics of final states with
high energy leptons.
Additionally the specific ionisation $dE/dx$ will be measured
with a precision better than 5\% to allow particle identification
in hadronic jets.

To reach these goals readout planes with the finest possible
granularity are required.
They must provide a single point resolution of 100 -- 150~$\mu$m 
and their systematic distortions must be controlled to better 
than 10~$\mu$m over the whole radius of 1.6~m.
Enough charge amplification has to be provided by the system to
keep the signals well above the noise level of modern readout electronics.
At the same time the produced ions have to be suppressed intrinsically,
because active gating seems impossible in between a TESLA bunch spacing
of only 337~ns.
Finally one aims for a minimum of material in front of the calorimeters,
especially in the endcap region.

A conventional TPC using a multi wire plane for charge amplification
is limited by the $\vec{E} \times \vec{B}$ effects in the region
close to the wires.
In strong magnetic fields these effects result in a broadening of the
electron cloud and a worsening of the resolution.
Additionally, as the wires define a preferred direction, the
reconstructed hit location depends on the projected angle between
track and wires.
The readout is done via pads which detect the induced signals
from the wires.
This induced signal is broader than the cloud
of arriving electrons and limits the granularity of the TPC
as well as the double track resolution.

Using Gas Electron Multipliers (GEM)~\cite{sauli} as the charge
amplifying system could solve some of the drawbacks of wire planes.
A GEM consists of a polymer foil about 50~$\mu$m thick and metal
coated on both sides.
It is perforated by a high density of small holes with diameters
of typically 50~--~100~$\mu$m.
Voltages up to 500~V between the two conducting sides generate
high electric fields in the order of 80~kV/cm inside the holes
where the gas amplification of the drifting electrons occurs.
When using GEM structures for the TPC end plate,
the pads directly detect the amplified electron cloud
which results in a fast and narrow charge signal.
Also the slow ion tail is cut off since the ion cloud does not
reach the induction region.
A GEM foil shows no preferred direction,
thus any $\vec{E} \times \vec{B}$ effects will be isotropic.
And finally, using highly different electric fields in front 
and behind the GEM, the back drift of ions produced inside 
the GEM holes ({\it ion feedback}) can be largely suppressed.

To demonstrate the advantages of a TPC with GEM readout a 
prototype chamber will be built within the R\&D activities of 
the linear collider TPC group~\cite{tpc-prc}.
It will be used to investigate wether the momentum accuracy, 
the double track resolution and the quality of the 
$dE/dx$ measurement as demanded for a detector at TESLA
can be reached.
The end plates of this prototype will contain up to three planes
of charge amplifying GEM foils.
The operation conditions of this multi GEM structure should be optimised
beforehand, because in the case of three GEM planes six
electric fields have to be set.
Therefore, a small test stand was set up at the RWTH Aachen.

The optimisation of a TPC readout plane should be done with respect 
to three important parameters:

\begin{itemize}
  \item The {\it electron transparency}, $T_\mathrm{elec}$, i.e.\
        the fraction of primary ionisation that experiences multiplication,
        should be near to 100\%.
  \item The {\it effective gain}, $G_\mathrm{eff}$, i.~e.\
        the number of electrons reaching the anode pads per 
        primary electron, should be sufficiently high.
  \item The {\it ion feedback}, $B_\mathrm{ion}$, i.~e.\
        the fraction of ions that reach the cathode plane,
        should be at a minimum.
\end{itemize}

To study the charge transfer mechanism of GEM structures we first
introduce the following variables for a single GEM foil:

\begin{itemize}
\item The {\it collection efficiency}, $C^{-(+)}$, i.e.\
      the fraction of electrons (ions) collected into the GEM holes
      per number of electrons (ions) arriving.
\item The {\it gain}, $G$, i.e.\
      the factor by which the number of electrons is increased by gas
      multiplication inside the GEM holes
\item The {\it extraction efficiency}, $X^{-(+)}$, i.e.\
      the fraction of electrons (ions) extracted from the GEM holes into
      the drift volume per number of electrons (ions) produced in the holes.
\item The {\it secondary extraction efficiency}, $X^+_\mathrm{sec}$, i.e.\
      the fraction of ions extracted from the GEM holes into
      the drift volume per number of ions which had been collected into 
      the holes. 
      This is different from the primary extraction, $X^+$, as will become
      clear in Section~\ref{results}.
      
\end{itemize}

These {\it transfer coefficients} have been determined as a function of 
the electric and magnetic fields by measuring the various
currents that appear on the electrodes of the GEM structure. 
Numerical simulations of the electric field fluxes allowed us to
predict these variables and to compare the prediction with measurements.
Additionally a parametrisation of the simulation has been established which
can be used in future to optimize the working point of the GEM structure
with respect to ion feedback and transparency.

\section{The Test Chamber}
\label{testchamber}
\subsection{Mechanical Setup}

\begin{figure}[ht]
 \begin{center}
 \mbox{\includegraphics[width=\textwidth]{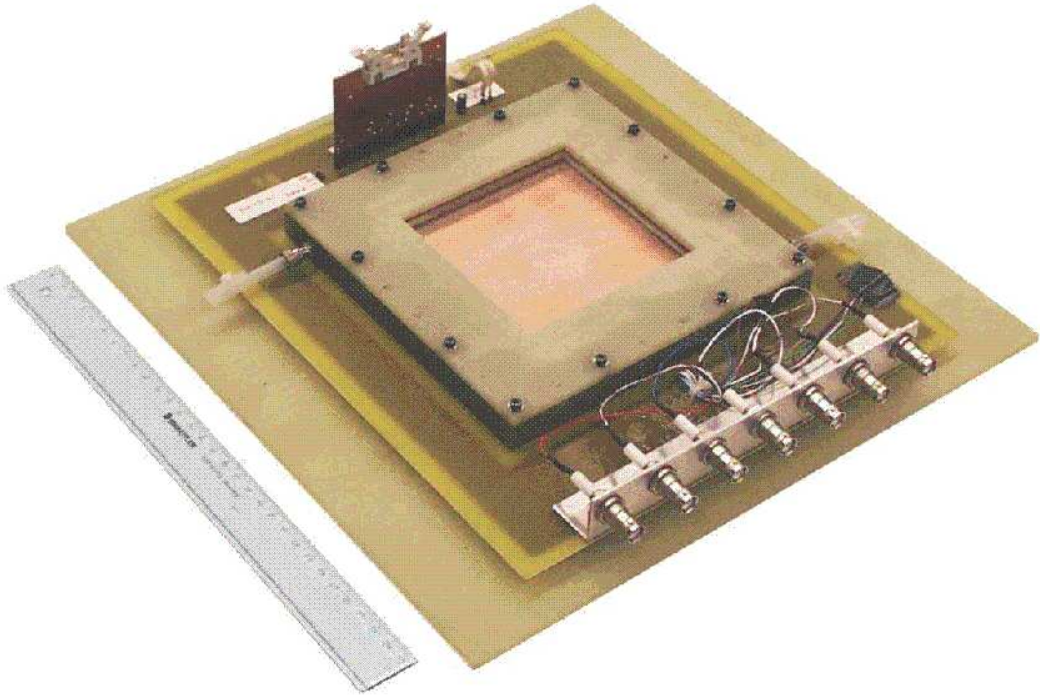}}
 \end{center}
 \caption{The test chamber used for the measurements}
 \label{fig_chamber}
\end{figure}

The test chamber is built on a PC board which is used as the anode plane. 
The sensitive area consists of a solid copper electrode of the same size 
as the GEM structures ($10 \times 10 \; \mathrm{cm}^2$).  
A bolt made from non conducting material is fixed in each
of the four corners 
This allows the mounting of up to three GEM foils glued on frames with 
corresponding holes.  The distances between the GEM foils (typically 2 mm) 
can be set by spacers.  On top of the last GEM the cathode, consisting 
of a fine metal mesh, is fixed in the same way as the GEMs.  
The gas volume is closed by a composite frame to the sides.  
On top it is covered by a 24 $\mu$m thick mylar foil which allows the 
photons from an ${}^{55}$Fe source to penetrate into the chamber.

The chamber has been operated alternatively with Ar/CO${}_2$ 82/18 
for the measurements without a magnetic field (Section~\ref{results}) 
or with Ar/CH${}_4$ 95/5 for the magnet tests at a rate of 2-6 l/h
(Section~\ref{magnetic}). 
It was irradiated from the top by an ${}^{55}$Fe source with an 
activity of the order of 100 MBq.

\subsection{Electric Setup}

\begin{figure}[ht]
 \begin{center}
 \mbox{\includegraphics[width=10cm]{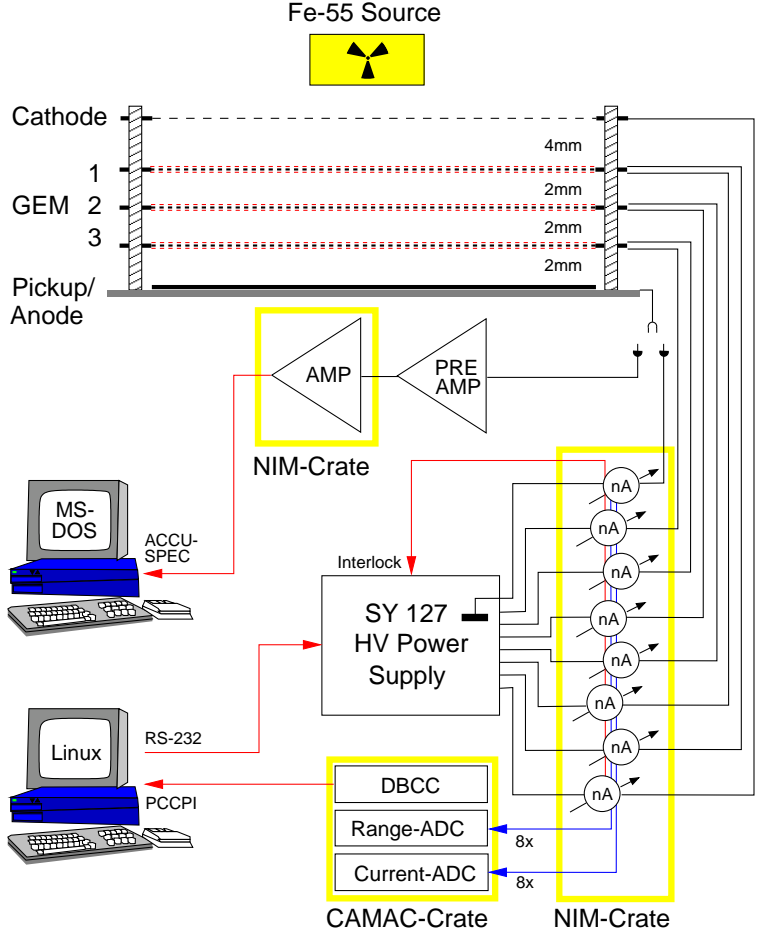}}
 \end{center}
 \caption{Experimental setup for the current measurements}
 \label{fig_setup}
\end{figure}

Each electrode (GEM surfaces and cathode) is connected to an individual channel
of a CAEN SY127 HV power supply, allowing the flexible setting of all electric 
fields and GEM voltages. 
Nano-Amperemeters \cite{beissel}, constructed in our electronics workshop 
and providing a resolution of approximately 0.1~nA, are inserted in the 
supply line of each HV channel. 
To measure the anode current, all pads and the surrounding copper are connected 
to ground via an individual nanoamperemeter. The high voltage control and current 
readout are handled by a custom application running on a Linux system.

\subsection{Analysis}

\begin{figure}[ht]
 \begin{center}
 \mbox{\includegraphics[width=10cm]{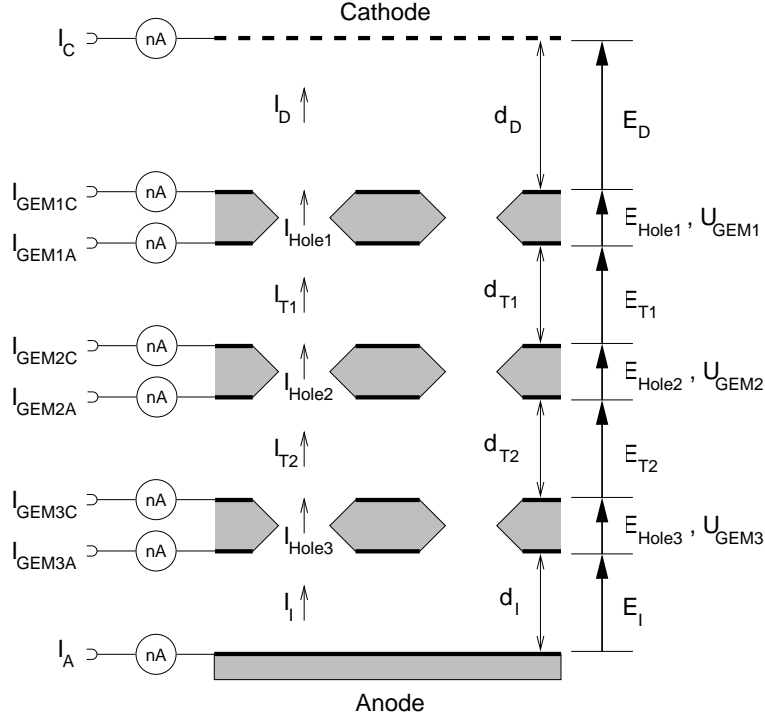}}
 \end{center}
 \caption{Assignment of fields and currents to the corresponding electrodes; 
   the volumes surrounding the GEMs are referred to as {\b d}rift, {\b t}ransfer, 
   and {\b i}nduction regions (from top to bottom). }
 \label{fig_bezeichnung}
\end{figure}

In Section~\ref{introduction}, the charge transfer coefficients are defined as 
ratios of charges. 
We use the corresponding currents, as charges are produced and transferred 
continuously. 
These currents of drifting electrons and Ions ($I_D$, $I_{T,n}$, $I_{Hole,n}$ 
and $I_I$ in Figure~\ref{fig_bezeichnung}) cannot be measured directly, but 
have to be calculated from the currents transferred on the electrodes 
($I_C$, $I_{GEM,n,C/A}$, $I_A$). 
The fact that every current has an electron and an ion component 
must be considered when calculating transfer coefficients for electrons or 
ions separately. 
Depending on the individual setup and coefficient, this can become quite complex. 
Details are described in~\cite{diplomarbeit_sven}. 

As an example, consider the electron extraction efficiency for GEM3. 
It is defined as the fraction of electrons extracted from the GEM holes into
the induction region per number of electrons produced in the holes, and can 
be calculated from the measured currents as

\begin{displaymath}
X^- = \frac{I_A}{I_A+I_{GEM3A}}~.
\end{displaymath}

This is a simple case, because in the region below GEM3 only electrons exist.

\section{Numerical Simulations and Parametrisations}
\label{simulations}
\begin{figure}[ht]
 \begin{center}
 \mbox{\includegraphics[width=0.8\textwidth]{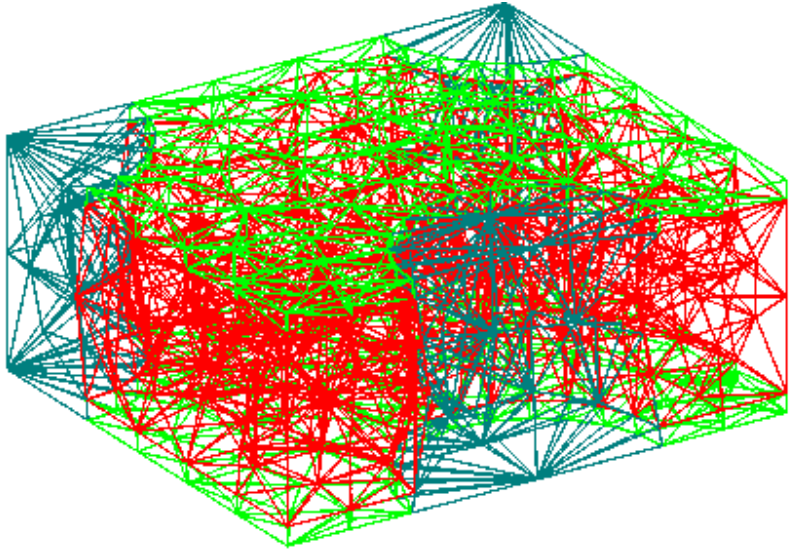}}
 \end{center}
 \caption{Coarse finite element mesh generated with Maxwell 3D}
 \label{fig_mesh}
\end{figure}

The numerical simulations are performed using
the finite element program Maxwell 3D of Ansoft Co.~\cite{maxwell}.
It allows to solve the Maxwell equations in three dimensional problems.
For the simulation of the electric fields inside and between the GEM
structures only the electrostatic problem solver is needed.

First the geometry of a unit cell of the GEM is constructed with the
CAD part of the program. 
Material properties are assigned to the geometrical objects using 
the provided database.  
Above and below the GEM foil 200~$\mu$m of material-free drift 
field are added to the model.  
Then electric potentials are assigned to each conducting plane.  
The boundary conditions at the border of the simulated unit cell are 
chosen such that the electric field is parallel to the border planes 
(Neumann boundaries).

Maxwell generates the finite element mesh by dividing the model 
volume into small tetrahedrons.  
The mesh can be refined in critical regions like sharp edges for example.  
The electric potential is calculated at each knot and each centre of the 
connection lines of the mesh by solving the Poisson equation numerically. 
Inside the tetrahedrons the potential is interpolated by a second order 
polynomial.

After a numerical solution has been found, it can be visualized using the
post processor.
For example, field vectors or potential lines can be drawn,
but also further calculations can be applied to the solution found.
Especially mean electric fields or the electric flux through
different areas is used extensively throughout this analysis.
For details of the simulation studies, see~\cite{diplomarbeit_blanka}.

\subsection{Electric field inside the GEM hole}

One of the most important parameters of a GEM structure is the mean
electric field inside the GEM hole, $E_\mathrm{hole}$.
We define this parameter from the integral of the electric field
over the area of the hole centre, $A$ ($\vec{n}$ is its normal vector):
\begin{equation}
\label{eqn_holefield}
E_\mathrm{hole} = \frac{\int_A \vec{E}\cdot\vec{n} \; d^2r}{\int_A \; d^2r}
\end{equation}

\begin{figure}[ht]
\begin{center}
\mbox{\includegraphics[width=\textwidth]{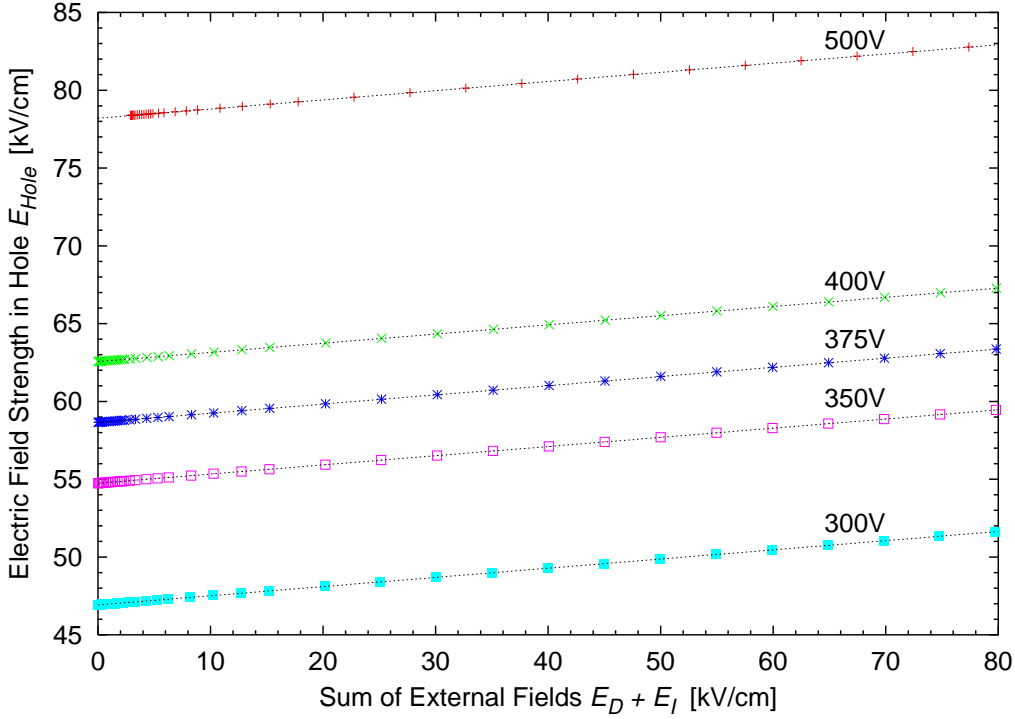}}
\end{center}
\caption{Electric field strength in GEM hole for several GEM voltages
 and different external fields}
\label{fig_holefield}
\end{figure}

The calculated field depends on the used GEM geometry, the voltage
across the GEM foil and the electric fields of the drift regions
above and below the GEM, as shown in Figure~\ref{fig_holefield}.
The plot suggests the following linear parametrisation
\begin{equation}
E_\mathrm{hole} = a \cdot U_\mathrm{GEM} + b \cdot (E_D + E_I) \; ,
\end{equation}
where $a$ and $b$ are parameters depending on the GEM geometry. 

\subsection{Electric field flux and charge transfer coefficients}

Neglecting diffusion effects across the GEM structure and
assuming zero magnetic field, all charges (electrons and ions)
will follow the electric field lines. 
It is also assumed that during gas multiplication charges are spread 
homogeneously across the complete cross section of the GEM hole. 
In this approximation it is possible to calculate the 
charge transfer coefficients from the electric field flux.

\begin{figure}[ht]
\begin{center}
\mbox{\includegraphics[height=0.50\textheight]{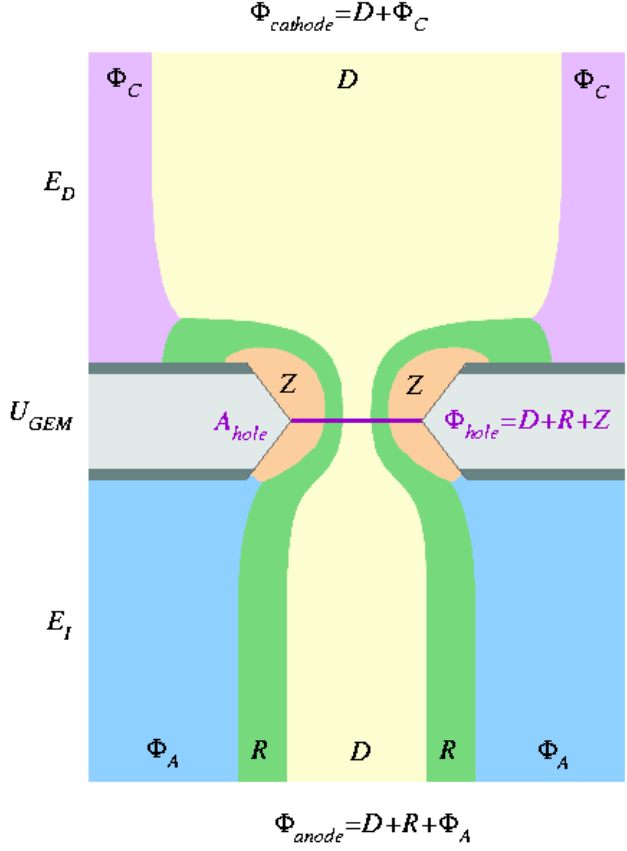}}
\end{center}
\caption{Model of the electric flux for one GEM hole in an
asymmetric field configuration. 
(Different electric fields above and below the GEM foil.)}
\label{fig_fluxsketch}
\end{figure}

Figure~\ref{fig_fluxsketch} shows a schematic plot of the electric flux 
through a GEM hole.
The total flux coming from the anode $\Phi_\mathrm{anode}$ 
and the flux going to the cathode $\Phi_\mathrm{cathode}$ can be
calculated from a surface integral over the electric field at the
bottom and top of the GEM model, respectively.
The total flux through the GEM hole is again the flux integral
over the area of the hole centre.

Furthermore, the electric flux can be separated into contributions 
from distinct regions illustrated by different shades in 
Figure~\ref{fig_fluxsketch}.
There are the fluxes $\Phi_\mathrm{C}$ and $\Phi_\mathrm{A}$ 
which go directly from the electrodes to the copper 
planes of the GEM.
The direction of the electric field changes between region
$\Phi_\mathrm{C}$ and $R$ and between $\Phi_\mathrm{A}$ and $Z$.
Therefore $\Phi_\mathrm{C}$ and $\Phi_\mathrm{A}$ can be extracted by a
numerical integration over that part of the copper plane,
where the electric field is pointing upwards.
The electric flux $D$ which goes from one electrode through the
GEM hole directly to the other electrode, the flux $R$ from the
anode through the hole onto the top side of the GEM surface and
the flux $Z$ going from the lower to the upper GEM surface 
cannot be calculated directly.

The collection efficiency $C$ is the fraction of the external flux
that is collected into the GEM hole:
\begin{equation}
\label{eqn_C}
C_\mathrm{top}    = \frac{D}{\Phi_\mathrm{cathode}} = 
\frac{\Phi_\mathrm{cathode}-\Phi_\mathrm{C}}{\Phi_\mathrm{cathode}},
\quad
C_\mathrm{bottom} = \frac{D+R}{\Phi_\mathrm{anode}} = 
\frac{\Phi_\mathrm{anode}-\Phi_\mathrm{A}}{\Phi_\mathrm{anode}} 
\end{equation}

The extraction efficiency $X$ can be calculated from the fraction
of the total electric flux originating from the hole and leaving 
the GEM structure towards cathode or anode:
\begin{equation}
\label{eqn_X}
X_\mathrm{top}    = \frac{D}{\Phi_\mathrm{hole}} = 
\frac{\Phi_\mathrm{cathode}-\Phi_\mathrm{C}}{\Phi_\mathrm{hole}},
\quad
X_\mathrm{bottom} = \frac{D+R}{\Phi_\mathrm{hole}} = 
\frac{\Phi_\mathrm{anode}-\Phi_\mathrm{A}}{\Phi_\mathrm{hole}} 
\end{equation}

\subsection{Simulation results and parametrisation}

\begin{figure}[ht]
\begin{center}
\mbox{\includegraphics[width=\textwidth]{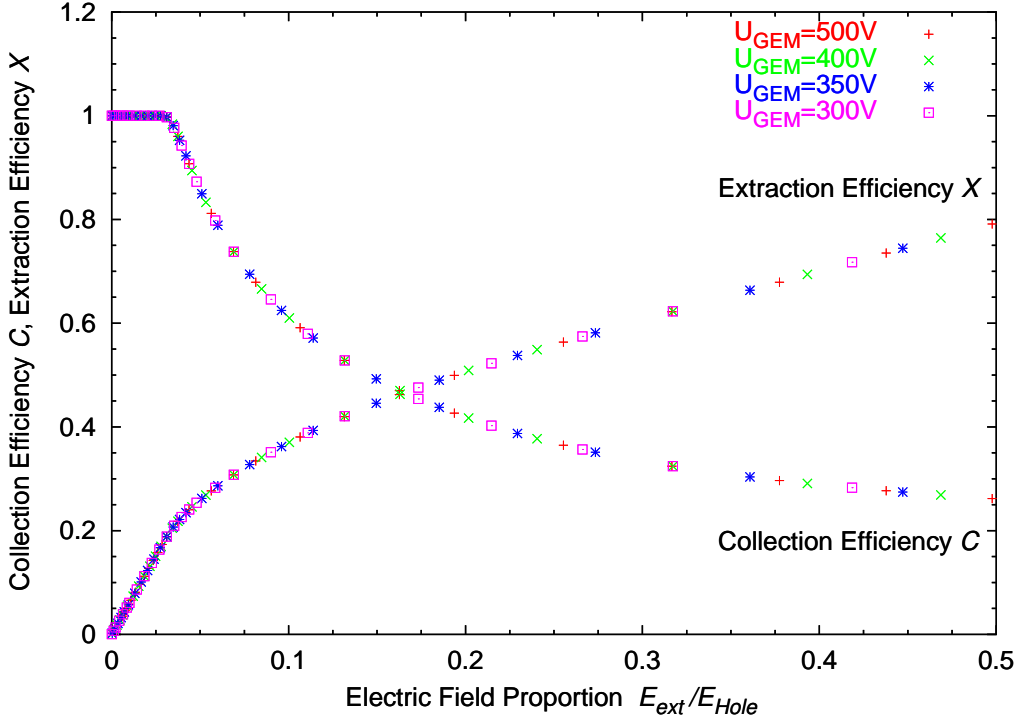}}
\end{center}    
\caption{Simulation of the extraction efficiency $X$ and collection efficiency $C$
 for several GEM voltages, $U_\mathrm{GEM}$ as a function of the ratio of
 external and hole field}
\label{fig_XC}
\end{figure}

Equations~\ref{eqn_C} and \ref{eqn_X} suggest that the collection efficiency
$C$ and extraction efficiency $X$ both vary with the ratio between the 
external flux and the flux through the GEM hole.
This is shown in Figure~\ref{fig_XC}, where the simulation results for
both coefficients are plotted versus the ratio $E_\mathrm{ext}/E_\mathrm{hole}$. 
Different markers show simulation results with different GEM voltages ranging 
from 300~V to 500~V.

From the field flux model also a relation between the collection and the 
extraction efficiency can be derived.
Let $\Phi_\mathrm{ext}$ be the flux onto an external electrode 
(anode or cathode) of size $A_\mathrm{ext}$.
As the electrode is far away from the GEM structure one
can assume a homogeneous electric field and the external
field strength is given by
\begin{equation}
  E_\mathrm{ext} = \frac{\Phi_\mathrm{ext}}{A_\mathrm{ext}} \; .
\end{equation}
On the other hand, using the mean electric field inside the 
GEM hole $E_\mathrm{hole}$ as defined in Formula~\ref{eqn_holefield} 
one can write the total electric flux through the GEM hole as
\begin{equation}
  \Phi_\mathrm{hole} = E_\mathrm{hole} \cdot A_\mathrm{hole} \; .
\end{equation}

Therefore, the ratio between extraction efficiency $X$ and 
collection efficiency $C$ is a linear function of the
ratio between external and hole field,
\begin{equation}
\label{eqn_fracXC}
  \frac{X}{C} = \frac{\Phi_\mathrm{ext}}{\Phi_\mathrm{hole}}
              = \frac{E_\mathrm{ext} \cdot A_{\mathrm{ext}}}
                     {E_\mathrm{hole} \cdot A_{\mathrm{hole}}}
              = \frac{1}{T_\mathrm{opt}} \cdot 
                \frac{E_\mathrm{ext}}{E_\mathrm{hole}} \; ,
\end{equation}
where the only geometrical parameter $T_\mathrm{opt}$ is the optical 
transparency of the GEM foil.

\begin{figure}[ht]
 \begin{center}
 \mbox{\includegraphics[width=\textwidth]{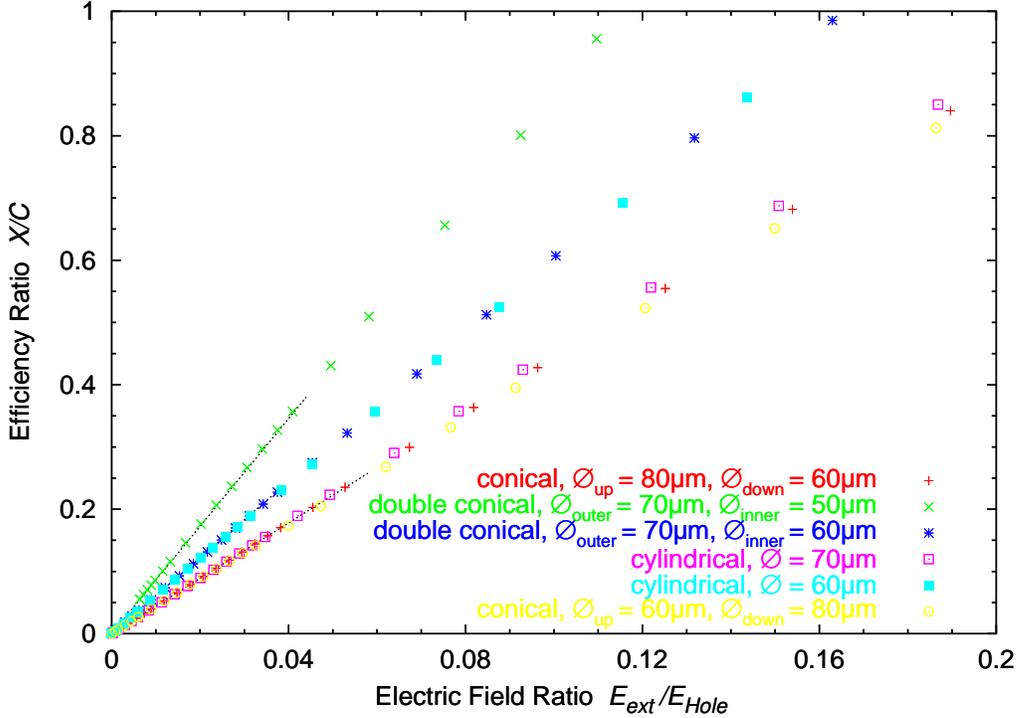}}
 \end{center}
 \caption{The ratio $X/C$ for different hole diameters and shapes.
  Conical GEM holes have an inner diameter smaller than the 
  outer diameter at the kapton surface
  and cylindrical holes have a constant diameter over
  the total depth of the hole.}
 \label{fig_ratioXC}
\end{figure}

Figure~\ref{fig_ratioXC} shows the predicted linear dependence between
the ratio $X/C$ and the field ratio $E_\mathrm{ext}/E_\mathrm{hole}$ 
calculated for different GEM geometries.
Six different geometries describing holes with different sizes and 
different shapes ({\it conical} and {\it cylindrical}) have been
simulated.
Geometries with equal optical transparency show the same slope.

\begin{figure}[ht]
 \begin{center}
 \mbox{\includegraphics[width=\textwidth]{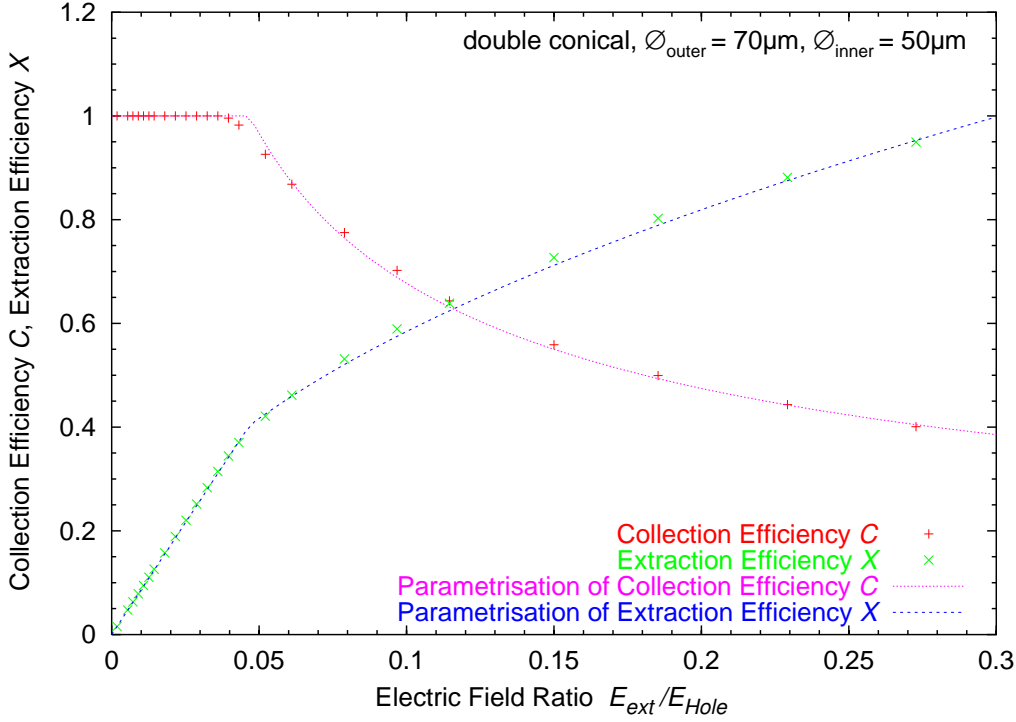}}
 \end{center}
 \caption{Simulated efficiencies $C$ and $X$ compared to their parametrisations}
 \label{fig_paraXC}
\end{figure}

The linear dependency between the logarithms of the collection efficiency
$C$ and the field ratio $E_\mathrm{ext}/E_\mathrm{hole}$ 
leads to a parametrisation of the simulated results using the 
following ansatz ($r,s\in\mathbb{R}^+$ are free parameters):
\begin{equation}
  C = \left\{ \begin{array}{ll} 
       1 & \qquad \mbox{for} \quad E_\mathrm{ext}/E_\mathrm{hole} \le r^{1/s} \\
       r \cdot (E_\mathrm{ext}/E_\mathrm{hole})^{-s}  
         & \qquad \mbox{for} \quad E_\mathrm{ext}/E_\mathrm{hole} > r^{1/s} \\
             \end{array} \right.
\end{equation}

Following Formula~\ref{eqn_fracXC} this gives in addition a parametrisation
of the extraction efficiency $X$:
\begin{equation}
  X = \left\{ \begin{array}{ll} 
       \frac{1}{T_\mathrm{opt}} \; (E_\mathrm{ext}/E_\mathrm{hole}) & \qquad \mbox{for} \quad E_\mathrm{ext}/E_\mathrm{hole} \le r^{1/s} \\
       \frac{r}{T_\mathrm{opt}}\;(E_\mathrm{ext}/E_\mathrm{hole})^{1-s}  
         & \qquad \mbox{for} \quad E_\mathrm{ext}/E_\mathrm{hole} > r^{1/s} \\
             \end{array} \right.
\end{equation}

Nearly perfect agreement between this parametrisation and the simulation
results over the entire range of the field ratio can be seen from
Figure~\ref{fig_paraXC}.

\section{Measurement Results and Comparison with Simulation}
\label{results}

\begin{figure}[ht!]
 \begin{center}
 \mbox{\includegraphics[width=\textwidth]{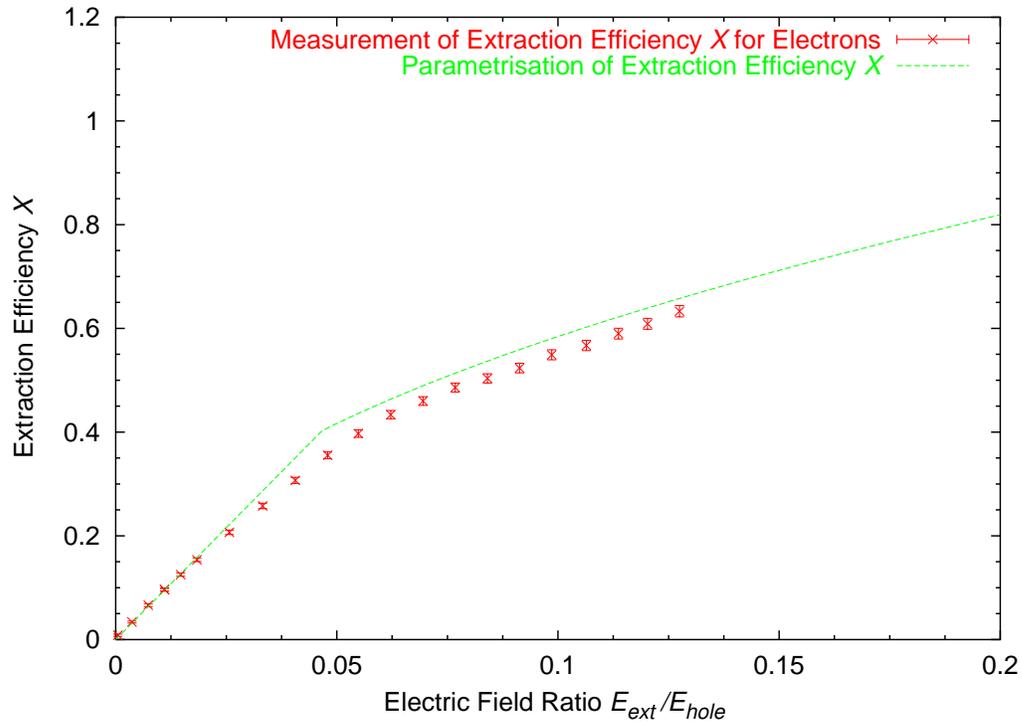}}
 \end{center}
 \caption{Measured extraction efficiency for electrons compared to simulation.}
 \label{fig_EX}
\end{figure}

\begin{figure}[hb!]
 \begin{center}
 \mbox{\includegraphics[width=\textwidth]{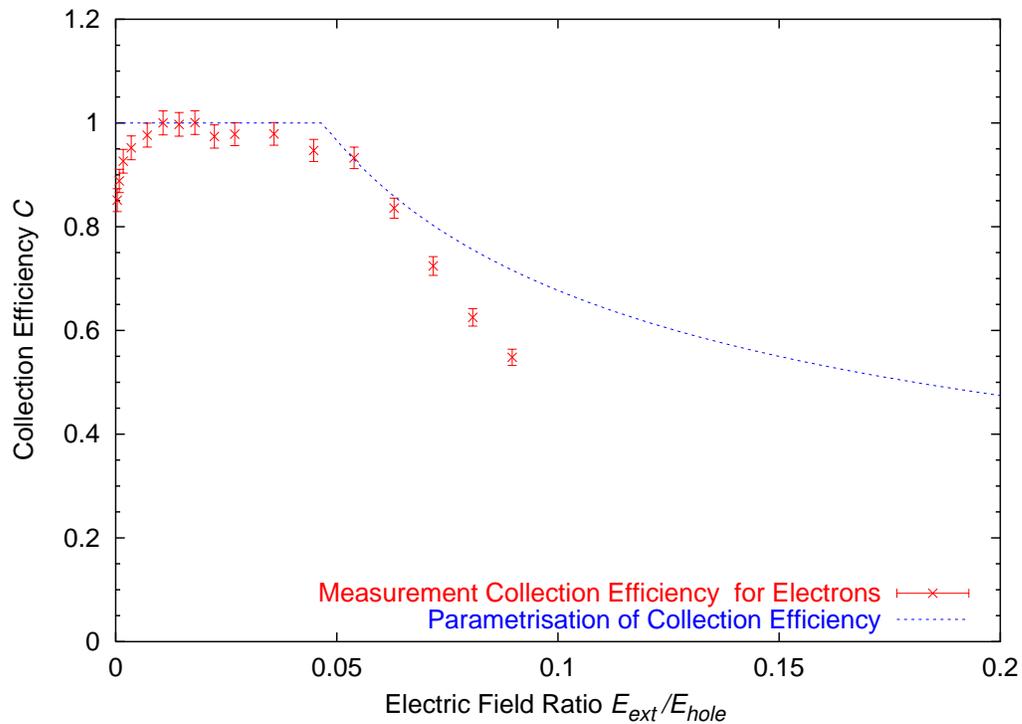}}
 \end{center}
 \caption{Measured collection efficiency for electrons compared to simulation.
	  As no absolute measurement was available the data points have been 
          scaled such that collection efficiency is 100\% at low field ratios 
          as suggested by the simulation.}
 \label{fig_EC}
\end{figure}

\begin{figure}[ht!]
 \begin{center}
 \mbox{\includegraphics[width=\textwidth]{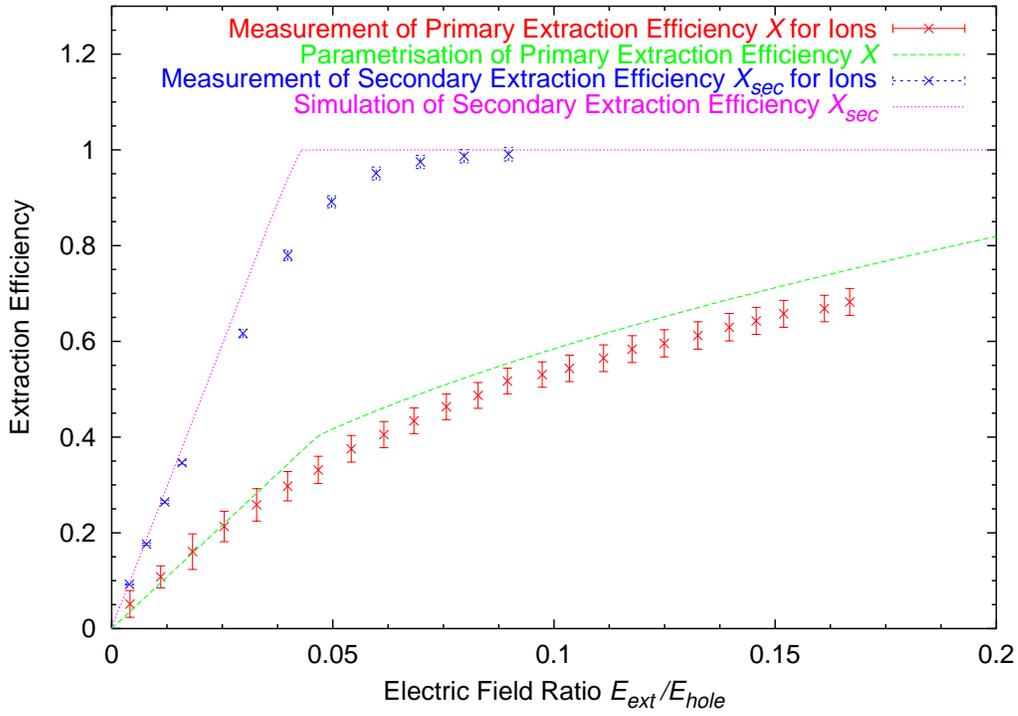}}
 \end{center}
 \caption{Measured primary and secondary extraction efficiency for ions compared to simulation.}
 \label{fig_IX}
\end{figure}

\begin{figure}[hb!]
 \begin{center}
 \mbox{\includegraphics[width=\textwidth]{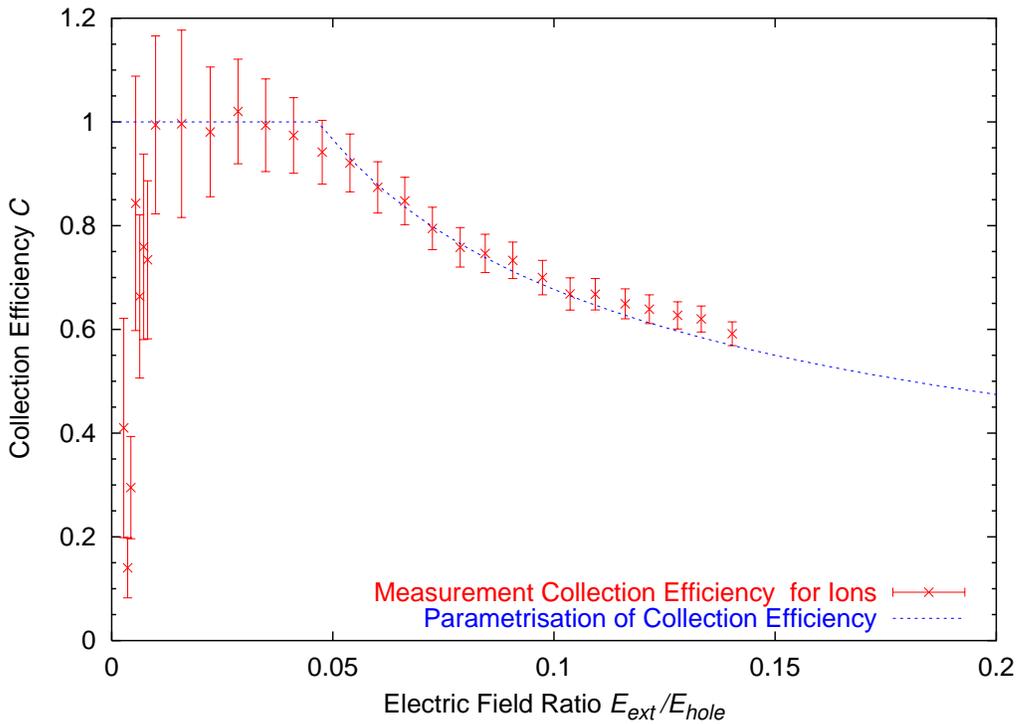}}
 \end{center}
 \caption{Measured collection efficiency for ions compared to simulation.}
 \label{fig_IC}
\end{figure}

Figures~\ref{fig_EX} through \ref{fig_IC} show the comparison between measurements made 
with the test chamber (Section~\ref{testchamber}) and the parametrisation obtained from 
the electrostatic simulations (Section~\ref{simulations}).

All data except for the electron collection efficiency follow the simulation. 
That shows that those coefficients are governed mostly by the electrostatic conditions 
in the chamber, not by gas effects such as diffusion. 
This assumption is also supported by the fact that the extraction efficiencies for 
electrons and ions are equal within their errors.

For the measured collection efficiency of both electrons and ions, there is a steep 
decrease for field ratios approaching $x=0$ while the simulation remains constant at $C=1$. 
We believe that this effect is due to recombination of charge pairs at very low drift 
velocities caused by the small electric field in the drift region.
In contrast, while showing a plateau of $C=1$ up to the same field ratio as the simulation, 
the collection efficiency for electrons decreases significantly steeper after that point. 
The origin of this effect is not yet understood.

The remaining difference between measurement and simulation for these variables is 
partly a consequence of the discrepancies between the ideal geometry of the GEM model 
in the simulations and the real GEMs. For example, changing the shape of the edges 
of the holes in the GEMs copper plating from perpendicular to the inclination of the 
conical holes in the polyimide causes a variation in the simulation results of the 
same order as the discussed differences.

The significant difference between primary and secondary ion extraction which does 
not occur for electrons, can be explained as follows:  
Through gas multiplication electrons are spread homogeneously across the cross section 
of every GEM hole they enter. 
Therefore a fraction of the electrons always occupies the radially outer regions of the 
hole penetrated by field lines ending on the top of the GEM and so decreasing 
the extraction efficiency. 
Primary ion extraction occurs through the same mechanism and shows the same values 
(compare Figures~\ref{fig_EX} and \ref{fig_IX}). 
However, ions subsequently collected into another GEM enter the hole in the middle, 
and because no multiplication and little diffusion occurs, they mostly remain on field 
lines continuing to the following GEM or electrode. 
This leads to the increased secondary extraction efficiency observed (Figure~\ref{fig_IX}). 
Understanding these processes, calculations from electrostatic simulations describe 
even the effect of secondary extraction by taking the electric flux entering 
the GEM hole into account (Figure~\ref{fig_fluxsketch}, \ref{fig_IX}).

\section{Effect of a Magnetic Field}
\label{magnetic}
Our primary motivation for making tests in a magnet was to investigate 
whether there is a significant drop in collection efficiency for GEMs 
in high B-fields parallel to the electric fields as suggested by the 
Langevin Formula

\begin{equation}
\vec{v}_{Drift}\propto\hat{\vec{E}}+\omega\tau\;(\hat{\vec{E}}\times\hat{\vec{B}})+\omega^2\tau^2\;(\hat{\vec{E}}\cdot\hat{\vec{B}})\;\hat{\vec{B}}~,
\end{equation}

where $\hat{\vec{E}}$ and $\hat{\vec{B}}$ are unit vectors of the fields 
and $\omega$ is the cyclotron frequency
\[
\omega=\frac{e}{m}B~.                                                                        
\]

The last term proportional to $B^2$, which gives the contribution 
along the magnetic field lines, could cause a drop in collection 
efficiency for high magnetic fields. When this term dominates, most 
electrons will no longer be collected into a GEM hole but will stay 
on drift lines perpendicular to the GEM surface and eventually reach 
the GEMs copper coating. 
Those charges will be lost for the signal and consequently decrease 
the chamber's $dE/dx$ capabilities due to the loss in primary 
ionisation statistics.

\subsection{Simulation}

\begin{figure}[ht!]
 \begin{center}
 \mbox{\includegraphics[width=\textwidth]{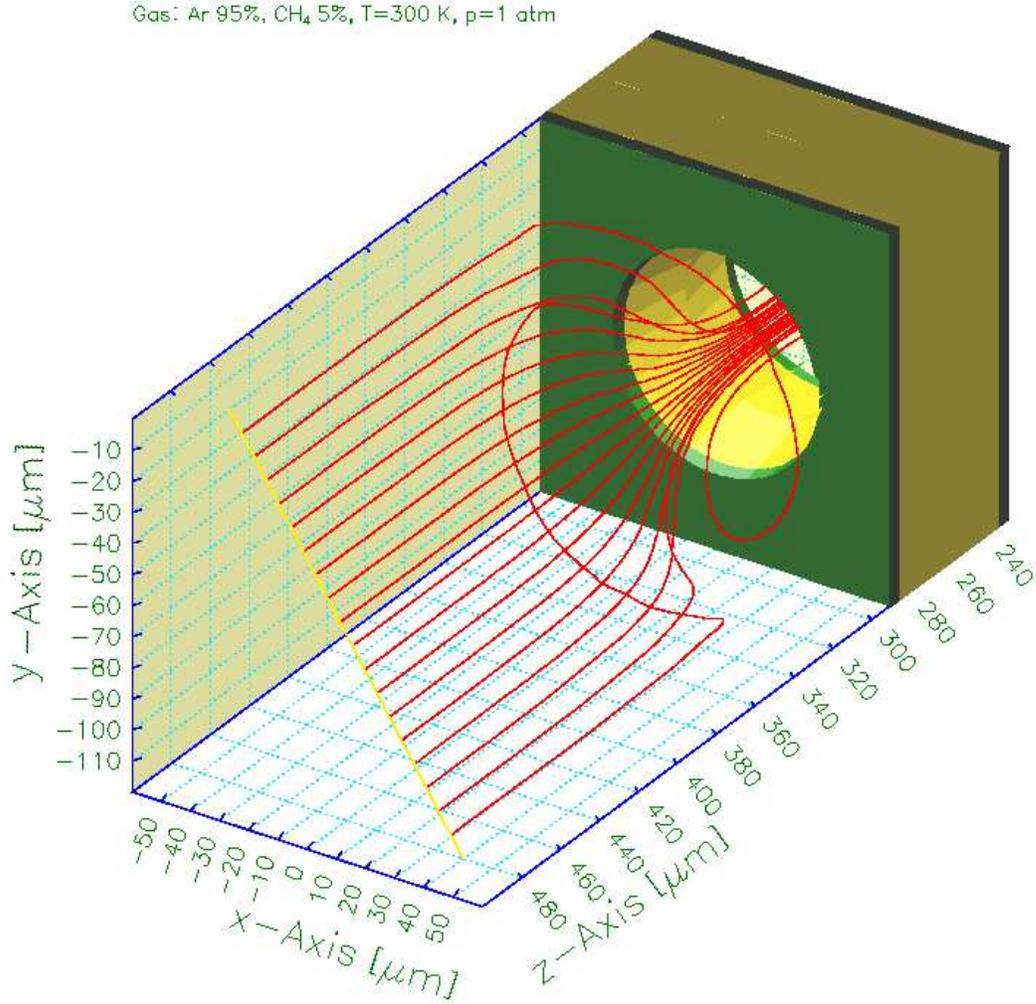}}
 \end{center}
 \caption{Garfield simulation of drift lines in a GEM hole in a 4~T magnetic field}
 \label{fig_garfield}
\end{figure}

To get a first qualitative view of the electrons behaviour, we calculated 
drift lines in the vicinity of a GEM foil pervaded by a 4~T magnetic field. 
This was done using the simulation program Garfield~\cite{garfield} which 
allows to take gas and magnetic field into account. 
We use a three dimensional electric field map of a GEM at 400~V with an 
external field of 200~V/cm obtained with Maxwell (see Section~\ref{simulations}). 
The gas mixture selected was Ar/CH${}_4$ 95/5 which is close to the mixture 
of Ar/CO${}_2$/CH${}_4$ 93/2/5 proposed in the TESLA Technical Design Report~\cite{tdr}. 
Each of the shown drift lines starts at a virtual straight ionisation track 
consisting of equally spaced electrons.

In Figure~\ref{fig_garfield} the effect of the Lorentz term
is well visible for electrons: they follow a track spiraling into the GEM hole.
In contrast there is no contribution to the drift velocity along the magnetic field lines.
Hence we have no indication for a significant drop in the
collection efficiency for primary electrons.

\subsection{Measurements}

For a quantitative analysis, we have conducted a series of current measurements 
with our test chamber in a non superconducting magnet limited to a maximum field of 2~T. 
All tests were performed with a magnetic field parallel to the electric fields in the chamber. 

For the test described here, we choose an electric field setup of 200~V/cm in 
the drift field, 5000~V/cm in the induction field and 2500~V/cm in the remaining fields. 
Each GEM is operated at a voltage of 310~V. 
The low drift field resembles the configuration in a TPC. 
The gas chosen is Ar/CH${}_4$ 95/5, the same as in the Garfield simulation.

\begin{figure}[ht]
 \begin{center}
 \mbox{\includegraphics[width=\textwidth]{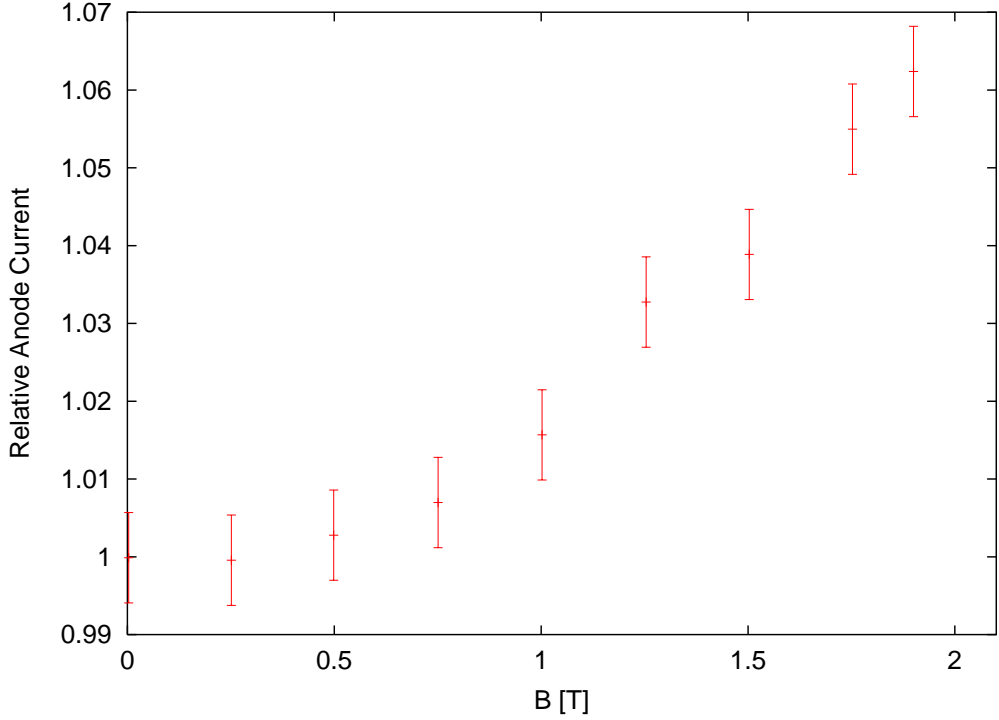}}
 \end{center}
 \caption{Measured anode current as a function of the magnetic field}
 \label{fig_currbfield}
\end{figure}

Opposite to the expected drop in collection efficiency, the signal current 
on the anode increases with $B$. 
At 2~T, it is approximately 6\% higher than without a magnetic field. 
Though surprising, this result is still compatible with a decreasing collection 
as anticipated but being dominated by a rise in extraction efficiency or gas gain. 

To find out about the mechanism for the observed rise in signal, another measurement was taken. 
The chamber was operated with all GEMs at 330~V and all fields at 1~kV/cm. 
This setting does not resemble the setting in a TPC (low drift field), 
but is chosen to allow simpler analysis of the charge transfer properties 
of the GEMs (same conditions at all GEMs, {\it periodic setup}): 
Because of the identical electrostatic conditions at all GEMs the corresponding 
transfer coefficients are equal and the signal current on the anode can be expressed as

\begin{eqnarray}
\nonumber I_A &= I_P&\cdot C_{GEM1}\cdot G_{GEM1}\cdot X_{GEM1}\\
\nonumber &&\cdot C_{GEM2}\cdot G_{GEM2}\cdot X_{GEM2}\\
\nonumber &&\cdot C_{GEM3}\cdot G_{GEM3}\cdot X_{GEM3} \\
&= &I_P\cdot C_{GEM}^3\cdot G_{GEM}^3\cdot X_{GEM}^3
\end{eqnarray}

with $I_P$ being the current of primary ionisation by the photons from the $^{55}$Fe source.

\begin{figure}[ht]
 \begin{center}
 \mbox{\includegraphics[width=\textwidth]{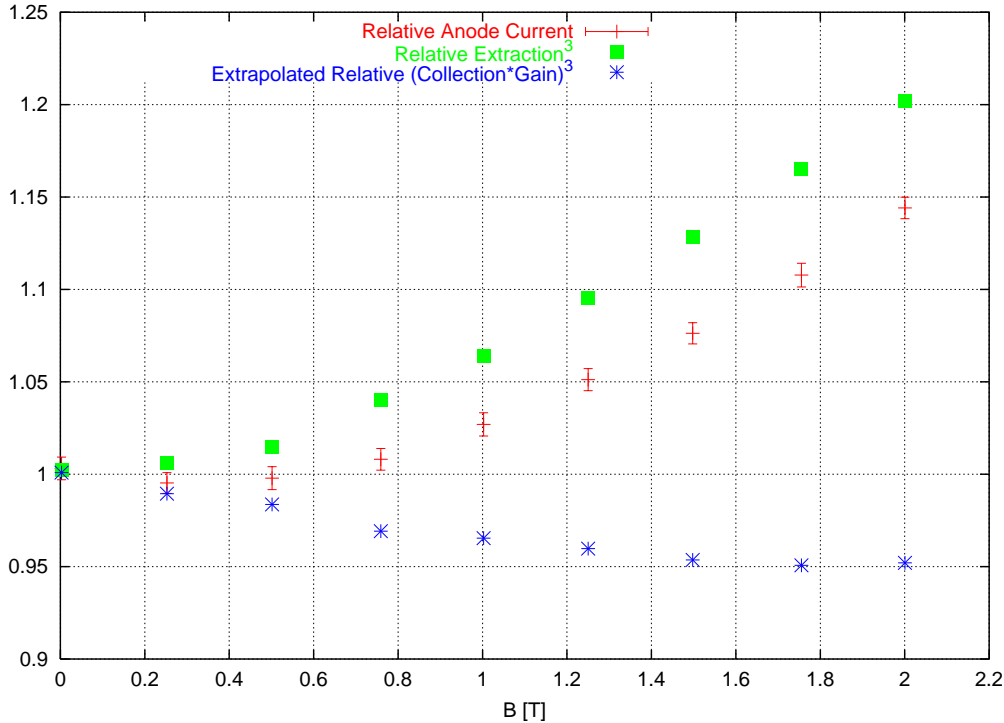}}
 \end{center}
 \caption{Anode current vs. magnetic field for the periodic setup}
 \label{fig_currbfield2}
\end{figure}

Figure~\ref{fig_currbfield2} shows the anode current normalised to its value at $B=0$. 
As before, it shows an increase with the magnetic field, this time by approximately 15\% at 2~T. 
So obviously the influence of a magnetic field on the signal current is dependent on 
the electrostatic chamber setup. 

From the currents in the chamber recorded during the measurement the extraction 
efficiency of GEM3, $X_{GEM3}$, was calculated. 
As discussed above, it is the same as for the two other GEMs, $X_{GEM3}=X_{GEM}$. 
The squares in Figure~\ref{fig_currbfield2} show the values of $X_{GEM}^3$, 
again normalised to their value at $B=0$. 
The increase is even greater than that of the anode current (approximately 20\%), 
which means that the product $C_{GEM}^3\cdot G_{GEM}^3$ must have decreased. 
Dividing the relative signal current by the calculated $X_{GEM}^3$ yields the 
value of $C_{GEM}^3\cdot G_{GEM}^3$, which is plotted as stars.

Obviously, the rise in the anode current is caused by an improved extraction 
efficiency for high magnetic fields. A drop in collection efficiency is hardly 
visible (note that for a single GEM, the effect is approximately $\sqrt[3]{95\%}\approx98\%$ 
which is not significant taking the error of the measurement into account).

This result proves that the operation of triple GEM structures in a 2~T magnetic field is possible; 
for higher fields, corresponding measurements are scheduled.

\section{Conclusions}
\label{conclusions}
It has been shown that the charge transfer coefficients of single GEM foils 
are predominantly determined by electrostatic conditions. 
Within the errors the results from measurements of the electric currents 
in the GEM structure are in agreement with simulations of the electric flux.
The charge transfer coefficients can be described by a set of simple 
parametrisations which match the simulation almost identically.

A difference has been found between the extraction efficiencies of ions 
out of GEM holes where they were produced by gas multiplication 
({\it primary extraction}) and out of GEM holes into which the ions were 
collected ({\it secondary extraction}). 
Electrons do not show this behavior, because they always experience 
multiplication in GEM holes (assuming a sufficiently large electric field) 
and are therefore distributed homogeneously across the volume of a GEM hole.

It has been demonstrated that for the working conditions in a TPC with
low external field (the TPC drift field) and high electric field
inside the GEM holes the ion feedback is intrinsically reduced to
the percent level keeping full transparency for primary electrons.
More sophisticated analysis of our parametrisation should allow
to find an optimal field configuration of the GEM structure.

Applying a magnetic field up to 2~T perpendicular to the surface 
of a triple GEM structure does not result in a decrease of the 
effective gain. 
Further studies are needed to understand the quantitative behaviour 
of the electron collection efficiency in magnetic fields up to 4~T.

\begin{ack}
We thank F.~Sauli and the CERN Gas Detector Development Group for supplying 
us with GEM foils and giving us a chance to look into the work of his team. 
We appreciate the support of T.~M\"uller's group at University of Karlsruhe, 
who manufactured the test chamber parts which helped us to get a quick start 
in conducting measurements. 
We are grateful to the IKP group of Forschungszentrum J\"ulich, 
especially O.~Felden and T.~Sagefka, for allowing us to use their test magnet.
We acknowledge the support of T.~Behnke, R.D.~Heuer and R.~Settles and the
congenial atmosphere in the Linear Collider TPC group.
Finally we thank S.~Bachmann fruitful discussions and the careful reading
of the manuscript.

\end{ack}

\end{document}